\documentclass[prl, twocolumn]{revtex4}
\usepackage{graphicx} 
\usepackage{dcolumn} 
\usepackage{multirow}
\begin{document}

\title{Metallic Indium Monolayers on Si(111)}

\author{Jae Whan Park}
\author{Myung Ho Kang}
\email{kang@postech.ac.kr}
\affiliation{Department of Physics, Pohang University of Science and Technology, Pohang 790-784, Korea}
\date{\today}

\begin{abstract}

Density-functional calculations are used to identify one-atom-thick metallic In overlayers on the Si(111) surface, which have long been sought in quest of the ultimate two-dimensional (2D) limit of free-electron-like metallic properties.
We predict two metastable single-layer In phases, one $\sqrt{7}\times\sqrt{3}$ phase with a coverage of 1.4 monolayer (ML; here 1 ML refers to one In atom per top Si atom) and the other $\sqrt{7}\times\sqrt{7}$ phase with 1.43 ML, which indeed match well with experimental evidences.
Both phases reveal quasi-1D arrangements of protruded In atoms, leading to 2D-metallic but anisotropic band structures and Fermi surfaces.  
This directional feature contrasts with the free-electron-like In-overlayer properties that are known to persist up to the double-layer thickness, implying that we may have achieved the 2D limit of free-electron-like In overlayers in previous studies of double-layer In phases.

\end{abstract}

\maketitle


How thin can metal films be yet retaining free-electron-like metallic properties \cite{fuch38,tell82}? 
It might be one atomic layer that represents the ultimate 2D limit of a crystalline film.
This fundamental question is in fact the very motivation underlying extensive experimental studies of the In/Si(111)-($\sqrt{7}\times\sqrt{3}$) surface \cite{kraf95, kraf97, hill99, rote03, sara06, zhan10, uchi11, yama11, uhm12, iwat13, yama13, uchi13, yosh14, shin15}, which has long been considered to represent one-atom-thick indium overlayers \cite{kraf95,kraf97}.
Fascinating 2D electronic features were reported, including the free-electron-like parabolic bands and circular Fermi surfaces \cite{rote03}, the persistence of superconductivity with a high T$_c$ close to the bulk value \cite{zhan10,uchi11}, and the intriguing metallic transport behavior \cite{yama11}, all of which have been referred to as revealing the ultimate 2D limit.


Unlike the expectations, however, the In/Si(111)-($\sqrt{7}\times\sqrt{3}$) surface was recently verified by  density-functional theory (DFT) calculations \cite{park12,park15} to actually represent two-atom-thick In overlayers, either rectangular (hereafter, $\sqrt{7}$-rect) or hexagonal ($\sqrt{7}$-hex), with 2.4 ML In coverage.
So far, there are two single-layer In phases with the coverage verified as 1.0 ML.
One is the 4$\times$1 phase, which is metallic but definitely one dimensional with weakly coupled In chains \cite{yeom99,bunk99,cho01}, and the other 2$\times$2 phase is known as an insulating 2D honeycomb lattice \cite{sara06,chou14,kwon14}. 
Thus, the single-layer limit of 2D-metallic In overlayers still remains to be explored.


Noteworthy in this regard is that there is $another$ In/Si(111)-($\sqrt{7}\times\sqrt{3}$) surface, differing from the verified $\sqrt{7}$-rect and $\sqrt{7}$-hex double-layer phases. 
This phase appears intermediately in between the 2$\times$2 and $\sqrt{7}$-rect phases when prepared by room-temperature (RT) In deposition onto the In/Si(111)-($\sqrt{3}\times\sqrt{3}$) surface and is known to transform into the honeycomb-like $\sqrt{7}\times\sqrt{7}$ phase during cooling down in the range from 265 to 225 K \cite{sara06}. 
This RT $\sqrt{7}\times\sqrt{3}$ phase was regarded as the $\sqrt{7}$-hex phase on the basis of similar STM images \cite{sara06}, but a recent DFT study clarified that it should be distinguished from the double-layer $\sqrt{7}$-hex phase \cite{park15}.
Moreover, in a latest STM study \cite{shin15}, the RT $\sqrt{7}\times\sqrt{3}$ phase was clearly identified as the so-called $striped$ phase, appearing as a minor phase coexisting with the 1.0-ML 4$\times$1 phase at high-temperature ($\sim$400$^{\circ}$C) preparations \cite{kraf97}.  
In microscopy studies, the striped $\sqrt{7}\times\sqrt{3}$ phase (hereafter, $\sqrt{7}$-stripe) appears 0.5 {\AA} higher than the single-layer 4$\times$1 phase \cite{iwat13} but substantially lower by 1.9 {\AA} than the double-layer $\sqrt{7}$-hex phase \cite{shin15}. 
This suggests that the $\sqrt{7}$-stripe phase is possibly one atom thick, but its actual In coverage and structure are not known.


In this paper, we use DFT calculations to identify single-layer metallic In phases on Si(111).
Our formation-energy calculations predict two metastable In phases, one 1.4-ML $\sqrt{7}\times\sqrt{3}$ phase and the other 1.43-ML $\sqrt{7}\times\sqrt{7}$ phase, which agree well with the aforementioned $\sqrt{7}$-stripe and $\sqrt{7}\times\sqrt{7}$ surfaces, respectively. 
Both phases reveal interesting quasi-1D structural features with protruded In atoms, leading to anisotropic 2D-metallic band structures.
Their electronic nature will be compared with those of the established double-layer In phases.


We perform DFT calculations by using the Vienna $ab$ $initio$ simulation package \cite{kres96} within the Perdew-Burke-Ernzerhof generalized gradient approximation \cite{perd96} and the projector augmented wave method \cite{bloc94}. 
The Si(111) surface is modeled by a periodic slab geometry with a slab thickness of 6 atomic layers and a vacuum spacing of about 12 {\AA}.
The calculated value 2.370 {\AA} is used as the bulk Si-Si bond length.
Indium atoms are adsorbed on the top of the slab, and the bottom is passivated by H atoms.
We use a plane-wave basis set of 246 eV and a 4$\times$6$\times$1 $k$-point mesh for the $\sqrt{7}\times\sqrt{3}$ unit cell.
All atoms but the bottom two Si layers are relaxed until the residual force components are within 0.01 eV/{\AA}.
Similar calculation schemes were successfully used in our In/Si(111) studies \cite{park12,park15}. 


\begin{figure}
\centering{ \includegraphics[width=8.0 cm]{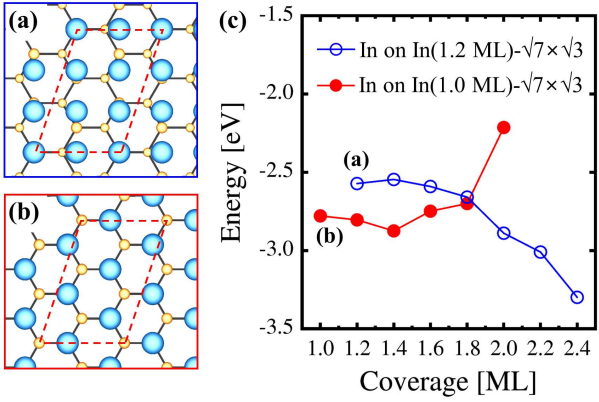} }
\caption{ \label{fig1}
(Color online)
Formation energy as a function of In coverage obtained from different initial substrates: 
(a) 1.2-ML rectangular and (b) 1.0-ML hexagonal In/Si(111)-($\sqrt{7}\times\sqrt{3}$), where large (small) balls represent In (Si) atoms.
Additional In atoms are tried on the hollow sites of (a) and (b), and the energy of the most stable configuration in each coverage is shown in (c). 
}
\end{figure} 


We first examine the energy stability of In overlayers on Si(111)-($\sqrt{7}\times\sqrt{3}$) in the coverage range of 1.0--2.4 ML, by regarding the 1.0 ML of the 2$\times$2 phase and the 2.4 ML of the $\sqrt{7}$-rect phase as the lower and upper bounds, respectively.  
For a given In coverage, we try various In configurations and identify the lowest-energy structure by comparing their formation energies, defined by $E = E_{\rm In/Si} - E_{\rm Si} - N \times E_{\rm In}$, where $E_{\rm In/Si}$, $E_{\rm Si}$ and $E_{\rm In}$ are the total energies of the In/Si(111)-($\sqrt{7}\times\sqrt{3}$) surface, the Si(111)-($\sqrt{7}\times\sqrt{3}$) surface, and the bulk In atom, respectively, and $N$ is the number of In atoms per $\sqrt{7}\times\sqrt{3}$ unit cell.


Figure 1 shows two energy curves obtained by using different In-covered substrates.
We first began with the 1.2-ML rectangular In phase shown in Fig. 1(a), which could be a precursor layer to the saturated double-layer (i.e., 2.4-ML $\sqrt{7}$-rect) phase, and searched the lowest-energy structures with increasing In atoms one by one.
The resulting energy curve (denoted by open circles) is getting lower with In coverage until arriving at the 2.4-ML $\sqrt{7}$-rect phase, which is known as a thermodynamically stable phase \cite{park12}.
The other starting configuration was the 1.0-ML hexagonal In phase shown in Fig. 1(b), which is also considerable as a precursor layer perfectly matching the Si(111)-(1$\times$1) surface.  
The resulting energy curve (filled circles) shows an interesting coverage dependence.
Whereas the final 2.0-ML double-layer phase appears far unstable, the lower-coverage phases are relatively stable with lower formation energies than the above-mentioned 1.2-ML series.
Moreover, there is a unique local-energy minimum at 1.4 ML, implying a metastable In phase.
This 1.4-ML phase has a relatively low formation energy (0.05 eV lower than the 1.0-ML 2$\times$2 phase and 0.16 eV higher than the 1.0-ML 4$\times$1 phase per $\sqrt{7}\times\sqrt{3}$ unit cell) and thus becomes a candidate for the $\sqrt{7}$-stripe phase that usually appears together with the 4$\times$1 or 2$\times$2 phase in experiments \cite{kraf97,sara06}.


\begin{figure}
\centering{ \includegraphics[width=8.0 cm]{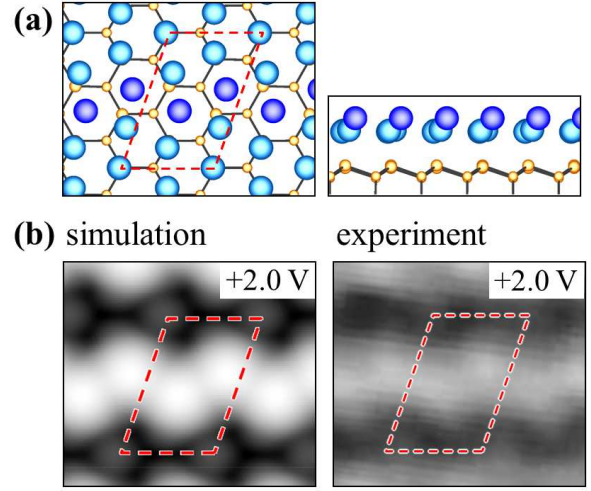} }
\caption{ \label{fig2}
(Color online)
1.4-ML In/Si(111)-($\sqrt{7}\times\sqrt{3}$).
(a) Atomic structure and (b) Simulated STM image, representing the surface of constant density with $\rho$=1$\times$$10^{-3}$ e/${\rm {\AA}}^3$. The experimental image was taken from Ref. \cite{sara06}.
}
\end{figure}       
                     

Figure 2 shows the atomic structure of the 1.4-ML phase.
Of the seven In atoms per $\sqrt{7}\times\sqrt{3}$ unit cell, five (denoted by light blue) strongly interact with the top Si atoms with an average interlayer spacing of 2.66 {\AA}, whereas the rest two (dark blue) are a little more protruded by 0.91 {\AA}.
The In coverage 1.4 ML is higher than 1.19 ML of the In(001) single layer but far lower than 2.4 ML of the double-layer $\sqrt{7}\times\sqrt{3}$ structures, and the layer puckering of 0.91 {\AA} is much smaller than the interlayer spacing of 2.40 {\AA} of the double-layer structures \cite{park12,park15}. 
Thus, the 1.4-ML In phase may well be regarded as a dense single layer.


The 1.4-ML phase indeed accounts well for the microscopic features of the $\sqrt{7}$-stripe surface \cite{kraf97,sara06,shin15}.
As seen in Fig. 2(b), its STM simulation compares well with the reported STM image of the $\sqrt{7}$-stripe surface \cite{sara06}: The protruded In atoms form a bright zigzag pattern along the $\sqrt{3}$ direction in good agreement with the experimental stripe image.
The 1.4-ML phase is also compatible with the reported topographic heights of the $\sqrt{7}$-stripe surface.
The height difference of 0.5 {\AA} between the $\sqrt{7}$-stripe and 4$\times$1 surfaces, measured by atomic force microscopy \cite{iwat13}, is close to our calculation of 0.36 {\AA} from the atomic heights of the 1.4-ML and  4$\times$1 phases.
The STM height difference of 1.9 {\AA} between the $\sqrt{7}$-stripe and $\sqrt{7}$-hex surfaces \cite{shin15} also compares well with our calculations of 1.54 {\AA} in atomic structure and 1.89 {\AA} in STM topograph.
Hereafter, we refer to the 1.4-ML phase as the $\sqrt{7}$-stripe phase.


\begin{figure}[t]
\centering{ \includegraphics[width=8.0 cm]{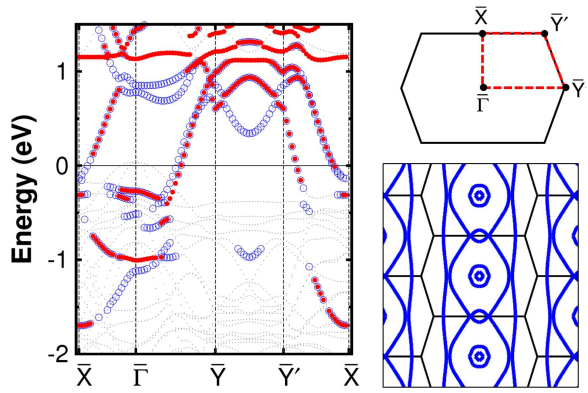} }
\caption{ \label{fig3}
(Color online)
Band structure of the $\sqrt{7}$-stripe phase.
Filled (open) circles represent In-derived states containing more than 20\% of charge in the protruded In atoms (more than 40\% in the other In atoms).
The $\sqrt{7}\times\sqrt{3}$ Brillouin zone and the calculated Fermi contours are shown in the right panel.
} 
\end{figure}


Figure 3 shows the electronic structure of the $\sqrt{7}$-stripe phase.
This single-layer In phase is 2D metallic with multiple bands crossing the Fermi level, but its 2D band structure is certainly anisotropic, well reflecting its quasi-1D structural character: 
It features two noticeable metallic bands with large dispersion along both $\overline{\Gamma}$-to-$\overline{Y}$ and $\overline{Y}^{\prime}$-to-$\overline{X}$ lines (parallel to the zigzag In-chain direction), whereas the metallic nature is much weaker along the $\overline{\Gamma}$-to-$\overline{X}$ and $\overline{Y}$-to-$\overline{Y}^{\prime}$ lines. 
The band gap along the $\overline{Y}$-to-$\overline{Y}^{\prime}$ line is well reflected in the anisotropic Fermi contours. 
This quasi-1D electronic nature is in contrast with the 2D free-electron-like features of the double-layer ($\sqrt{7}$-rect and $\sqrt{7}$-hex) phases \cite{park12,park15}.


\begin{figure} 
\centering{ \includegraphics[width=6.0 cm]{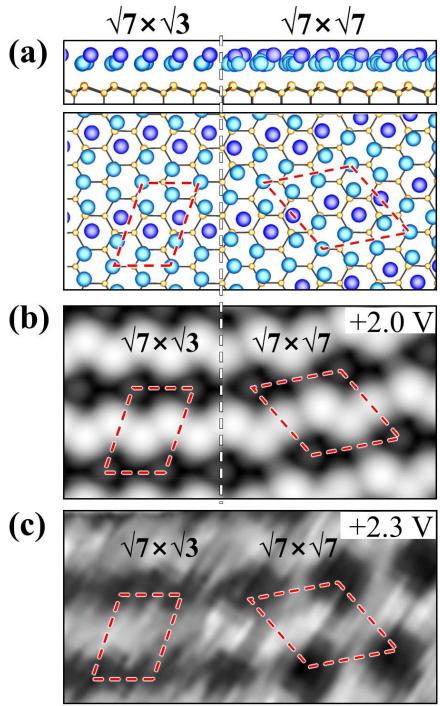} }
\caption{ \label{fig4}
(Color online)
Phase boundary between the 1.4-ML $\sqrt{7}$-stripe and 1.43-ML $\sqrt{7}\times\sqrt{7}$ phases:
(a) Atomic structure, (b) Simulated STM image, and (c) Experimental STM image taken from Ref. \cite{sara06}.
}
\end{figure}


It is interesting to further identify the $\sqrt{7}\times\sqrt{7}$ phase that is transformed from the $\sqrt{7}$-stripe phase at low temperatures (LTs) of 225--265 K \cite{sara06}.
We extended our formation-energy calculations for the $\sqrt{7}\times\sqrt{7}$ unit cell in the coverage range of 1.0--2.0 ML and found a unique local-energy minimum at 1.43 ML (corresponding to 10 In atoms per $\sqrt{7}\times\sqrt{7}$ unit cell).
As seen in Fig. 4, this 1.43-ML $\sqrt{7}\times\sqrt{7}$ phase also retains a quasi-1D structural nature with protruded In atoms, similar to the $\sqrt{7}$-stripe phase but with a different chain direction.
Fascinating is its simulated STM image that reproduces well not only the experimental honeycomb-like image of the LT $\sqrt{7}\times\sqrt{7}$ phase but also the sharp phase boundary with the coexisting $\sqrt{7}$-stripe phase \cite{sara06}.
We readily identify the 1.43-ML phase as the LT $\sqrt{7}\times\sqrt{7}$ phase.
This assignment is also sound energetically: 
The 1.43-ML phase has a lower formation energy (by 0.10 eV per $\sqrt{7}\times\sqrt{3}$ unit cell) than the 1.4-ML $\sqrt{7}$-stripe phase, accounting well for the preference of the $\sqrt{7}\times\sqrt{7}$ phase at low temperatures.
At elevated temperatures, however, the 1.4-ML $\sqrt{7}$-stripe phase would be favorable with a slightly lower In coverage because of thermal expansions. 


\begin{figure}[b] 
\centering{ \includegraphics[width=8.0 cm]{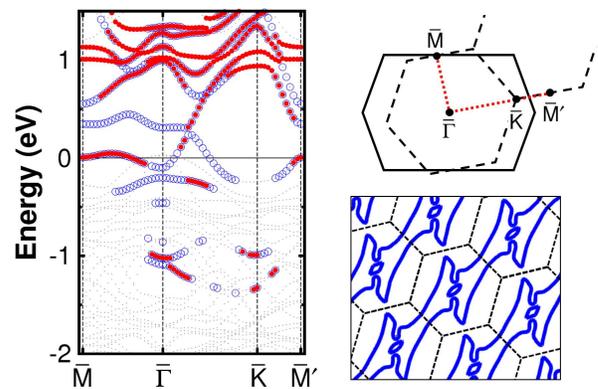} }
\caption{ \label{fig5}
(Color online)
Band structure of the $\sqrt{7}\times\sqrt{7}$ phase. 
The $\sqrt{7}\times\sqrt{7}$ Brillouin zone (dashed lines) and the calculated Fermi contours are shown in the right panel.
}
\end{figure}


Figure 5 shows the electronic structure of the $\sqrt{7}\times\sqrt{7}$ phase.
The 2D-metallic band structure is also anisotropic with dispersive metallic bands along the $\overline{\Gamma}$-to-$\overline{K}$ line and a non-dispersive metallic band along the $\overline{\Gamma}$-to-$\overline{M}$ direction, and so are the Fermi contours.


\begin{figure} 
\centering{ \includegraphics[width=7.0 cm]{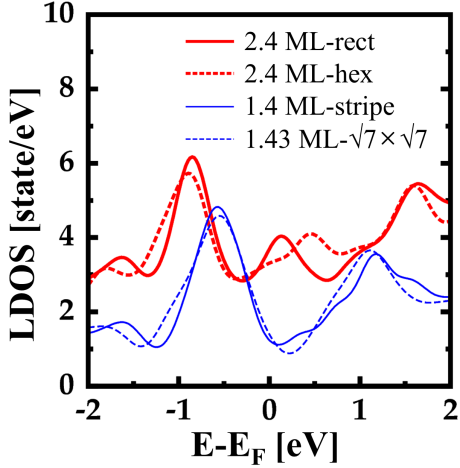} }
\caption{ \label{fig6}
(Color online)
Layer-resolved LDOS for the single- and double-layer indium phases. 
For comparison, the LDOS of 1.43 ML-$\sqrt{7}\times\sqrt{7}$ is normalized to the $\sqrt{7}\times\sqrt{3}$ unit cell.
}
\end{figure}


Figure 6 shows the In-derived local density of states (LDOS) of the $\sqrt{7}$-stripe and $\sqrt{7}\times\sqrt{7}$ phases.
Both phases reveal almost the same LDOS spectra, well reflecting their similar In coverages and quasi-1D structural nature.
Their LDOS spectra, however, differ clearly from those of the double-layer phases. 
The main peaks located at about $-$0.55 eV are much higher in energy than the peaks of the double-layer phases (at about $-$0.90 eV), indicating that the In-derived states are not yet sufficiently stabilized in the single-layer phases.
It is also noticeable that the single-layer phases have much weaker LDOS values at the Fermi level than the double-layer phases, implying that the free-electron-like nature of In overlayers is greatly suppressed in the single-layer regime by the dominant Si-In interactions.


In conclusion, the striped In/Si(111)-$\sqrt{7}\times\sqrt{3}$ phase and its low-temperature $\sqrt{7}\times\sqrt{7}$ phase have been verified to represent one-atom-thick metallic In overlayers.
Both phases, however, reveal quasi-1D structural features, and the resulting anisotropic 2D band structures  and Fermi surfaces contrast with those of the double-layer In phases that still retain the free-electron-like metallic properties \cite{park12,park15}. 
This strongly suggests that the ultimate 2D limit of free-electron-like In overlayers on Si(111) could be In double layers. 
At least one buffer layer may be needed to screen the rather strong substrate interactions, exactly as the same that the Dirac cone of graphene is realized not in monolayers but in bilayers or by proper intercalations, when grown on interactive substrates such as SiC(001) \cite{ohta06, varc07} and Ni(111) \cite{vary08,kang10}.


This work was supported by the National Research Foundation of Korea (Grant No. 2011-0008907). 


\newcommand{\PR}[3]{Phys.\ Rev.\ {\bf #1}, #2 (#3)}
\newcommand{\PRL}[3]{Phys.\ Rev.\ Lett.\ {\bf #1}, #2 (#3)}
\newcommand{\PRA}[3]{Phys.\ Rev.\ A\ {\bf #1}, #2 (#3)}
\newcommand{\PRB}[3]{Phys.\ Rev.\ B\ {\bf #1}, #2 (#3)}
\newcommand{\RMP}[3]{Rev.\ Mod.\ Phys.\ {\bf #1}, #2 (#3)}
\newcommand{\PST}[3]{Phys.\ Scr.\ T\ {\bf #1}, #2 (#3)}
\newcommand{\PML}[3]{Phil.\ Mag.\ Lett.\ {\bf #1}, #2 (#3)}
\newcommand{\SCI}[3]{Science\ {\bf #1}, #2 (#3)}
\newcommand{\SSA}[3]{Surf.\ Sci.\ {\bf #1}, #2 (#3)}
\newcommand{\SSCO}[3]{Solid\ State\ Comm.\ {\bf #1}, #2 (#3)}
\newcommand{\SSR}[3]{Surf.\ Sci.\ Rep.\ {\bf #1}, #2 (#3)}
\newcommand{\SRL}[3]{Surf.\ Rev.\ Lett.\ {\bf #1}, #2 (#3)}
\newcommand{\NA}[3]{Nature\ {\bf #1}, #2 (#3)}
\newcommand{\NAT}[3]{Nat.\ Phys.\ {\bf #1}, #2 (#3)}
\newcommand{\NAM}[3]{Nature\ Mater.\ {\bf #1}, #2 (#3)}
\newcommand{\JP}[3]{J.\ Phys.\ {\bf #1}, #2 (#3)}
\newcommand{\JACS}[3]{J.\ Am.\ Chem.\ Soc.\ {\bf #1}, #2 (#3)}
\newcommand{\JAP}[3]{J.\ Appl.\ Phys.\ {\bf #1}, #2 (#3)}
\newcommand{\JCP}[3]{J.\ Chem.\ Phys.\ {\bf #1}, #2 (#3)}
\newcommand{\JKPS}[3]{J.\ Korean Phys.\ Soc.\ {\bf #1}, #2 (#3)}
\newcommand{\JPCS}[3]{J.\ Phys.\ Chem.\ Solids.\ {\bf #1}, #2 (#3)}
\newcommand{\JVSA}[3]{J.\ Vac.\ Sci.\ Technol.\ A\ {\bf #1}, #2 (#3)}
\newcommand{\JJAP}[3]{Jpn.\ J.\ Appl.\ Phys.\ {\bf #1}, #2 (#3)}
\newcommand{\ASS}[3]{Appl.\ Surf.\ Sci.\ {\bf #1}, #2 (#3)}
\newcommand{\APL}[3]{Appl.\ Phys.\ Lett.\ {\bf #1}, #2 (#3)}
\newcommand{\APE}[3]{Appl.\ Phys.\ Exp.\ {\bf #1}, #2 (#3)}
\newcommand{\BBPC}[3]{Ber.\ Bunsen-Ges.\ Phys.\ Chem.\ {\bf #1}, #2 (#3)}
\newcommand{\CPL}[3]{Chem.\ Phys.\ Lett.\ {\bf #1}, #2 (#3)}
\newcommand{\LTP}[3]{Low\ Temp.\ Phys.\ {\bf #1}, #2 (#3)}
\newcommand{\TSF}[3]{Thin\ Solid\ Filims\ {\bf #1}, #2 (#3)}
\newcommand{\VAC}[3]{Vacuum\ {\bf #1}, #2 (#3)}
\newcommand{\EL}[3]{Europhys.\ Lett.\ {\bf #1}, #2 (#3)}
\newcommand{\IJMPB}[3]{Int.\ J.\ Mod.\ Phys.\ B.\ {\bf #1}, #2 (#3)}
\newcommand{\CJCP}[3]{Chin.\ J.\ Chem.\ Phys.\ {\bf #1}, #2 (#3)}
\newcommand{\NRL}[3]{Nano.\ Res.\ Lett.\ {\bf #1}, #2 (#3)}
\newcommand{\PLA}[3]{Phys.\ Lett.\ A.\ {\bf #1}, #2 (#3)}
\newcommand{\NT}[3]{Nanotechnology {\bf #1}, #2 (#3)}
\newcommand{\PCS}[3]{Proc.\ Cambridge Philos.\ Soc. {\bf #1}, #2 (#3)}
\newcommand{\PNAS}[3]{Proc.\ Natl.\ Acad.\ Sci. {\bf #1}, #2 (#3)}
\newcommand{\NS}[3]{Nanoscale\ {\bf #1}, #2 (#3)}
\newcommand{\CR}[3]{Chem.\ Rev.\ {\bf #1}, #2 (#3)}

\end{document}